\documentclass[copyright]{eptcs}
\providecommand{\event}{UNIF 2010} 
\usepackage{breakurl}             

\usepackage[latin1]{inputenc}
\usepackage{epsfig, amssymb, amsmath}
\usepackage{latexsym, graphics, graphicx}
\usepackage{yfonts}
\usepackage{proof}
\usepackage{algorithmic}
\usepackage{algorithm}
\usepackage{amsthm} 
\usepackage{amsxtra}
\usepackage{epsfig}
\usepackage{color}
\usepackage{float}
\newcommand{\ignore}[1]{}

\newcommand{\E}{\mathcal{E}}

\newcommand{\eq}{\mathcal{EQ}}

\newcommand{\ueq}{=_{}^?}

\theoremstyle{plain}
\newtheorem{theorem}{Theorem}[section]
\newtheorem{Lemma}[theorem]{Lemma}

\theoremstyle{definition}
\newtheorem{definition}[theorem]{Definition}

\title{Unification modulo a partial theory of exponentiation}
\author{Deepak Kapur\footnote{Partially supported by the NSF grants
CNS-0831462 and CNS-0905222}
\institute{University of New Mexico\\
Department of Computer Science}
\email{kapur@cs.unm.edu}
\and
Andrew Marshall\footnote{Partially supported by the NSF grants
CNS-0831209 and CNS-0905286}
\institute{University at Albany--SUNY\\
Computer Science Department}
\email{\quad marshall@cs.albany.edu }
\and Paliath Narendran\footnote{Partially supported by the NSF grants
CNS-0831209 and CNS-0905286} 
\institute{University at Albany--SUNY\\
Computer Science Department}
\email{dran@cs.albany.edu}
}
\def\titlerunning{Unification modulo a partial theory of exponentiation}
\def\authorrunning{D. Kapur, A. Marshall \& P. Narendran}
\begin{document}
\maketitle

\begin{abstract}
Modular exponentiation is a common mathematical operation in
modern cryptography.  This, along with modular multiplication at
the base and exponent levels (to different moduli) plays an
important role in a large number of key agreement protocols. In
our earlier work~\cite{KNW03a, KNW03b} we gave many
decidability as well as undecidability results for multiple
equational theories, involving various properties of modular
exponentiation. Here, we consider a partial subtheory focussing
only on exponentiation and multiplication operators. Two main
results are proved. The first result is positive, namely, that
the unification problem for the above theory (in which no
additional property is assumed of the multiplication operators)
is decidable. The second result is negative: if we assume that
the two multiplication operators belong to two different abelian
groups, then the unification problem becomes undecidable. This
result is established using a construction patterned after those
employed in~\cite{KNW03a, Narendran93onthe}  by
reducing Hilbert's $10^{th}$ problem to the unification problem.
\end{abstract}

\section{Introduction}

With network use and online transactions becoming all pervasive in many
applications,
especially online shopping, social networking, video-conferencing, group
conferencing, and e-voting
etc, multi-party and group protocols need to be employed.
These protocols are often complex, rich and sophisticated, built as a
collection of
protocols, whose interaction is often quite complex.
Their reliability and
security thus become a critical issue, especially in case the protocols
use arithmetic operators, such as modular multiplication and exponentiation
and boolean operators such as {\em exclusive-or}~\cite{DBLP:journals/corr/abs-cs-0610014}.
In collaboration with the Maude-NPA
team~\cite{escobar-meadows-meseguer-secret}, 
we have developed an approach for analyzing whether a given
protocol is vulnerable to specific attacks by modeling the protocol
as a state machine and an execution of the protocol as a sequence of state
transitions.
The search space is explored using unification and narrowing techniques
to handle equational properties of the operators used in a protocol.

Modular exponentiation is a common mathematical operation in
modern cryptography. This, along with modular multiplication at the base
and exponent levels (to different moduli) plays an important role
in the El Gamal signature scheme, the Nyberg-Rueppel key agreement
protocol (Protocol~5.3
in~\cite{BoydMathuria}), and the
MTI and Yacobi-Shmuely
protocols for public
key distribution (Protocols~5.7 and 5.33 in~\cite{BoydMathuria}).
In our earlier work~\cite{KNW03a,KNW03b} we gave
many decidability as well as undecidability results for multiple equational
theories, involving various properties of modular exponentiation.
Here, consider a partial subtheory
focussing only on exponentiation and multiplication operators.

The axioms of the theory are
\vspace*{-.1in}
 \begin{eqnarray*}
exp(g(X), Y) & = & g(X \circledast Y)\\
exp(X*Y, Z) & = & exp(X, Z) * exp(Y, Z)
\end{eqnarray*}
\vspace*{-.25in}

Here $exp$ is the exponentiation operator
and $g$ is exponentiation over a {\em fixed base,\/} such
as $2_{}^n$. The multiplication operators $*$ and $\circledast$
are often modulo a prime $p$ and $p-1$, respectively. The reason
for modeling two different exponentiation operators is that in a 
large majority of protocols, many operations are done using a fixed 
base. In addition, when specifying such protocols in Maude, as in 
Maude-NPA, the use of the subsort mechanism can make unification more 
efficient if the first argument in $exp$ is fixed.

Two main results are proved. The first result is positive,
namely, that the unification problem for the above theory (in
which no additional property is assumed of the multiplication
operators) is decidable. The second result is negative: if we
assume that the multiplication operators $*$ and $\circledast$
belong to two different abelian groups, then the unification
problem becomes undecidable. This result is established using a
construction patterned after those employed in
\cite{KNW03a,Narendran93onthe} by reducing Hilbert's $10^{th}$
problem to the theory.

The decidability result uses a novel construction and is
discussed in the next three sections. The next section
models the equational properties of the above two axioms as an inference system.
Section~3 analyzes possible reasons when the unification fails, corresponding
to the function clashes, occur-check, and an infinite application of one of the 
inference rules.
Section~4 gives the unification algorithm along with a termination proof.
The final section sketches the undecidability proof for the equational theory in which,
along with the above two axioms, the multiplication operators come
from abelian groups.

\vspace{-0.1in}
\section{Inference Rules}
Below we present a set of inference rules for unification. Termination
of these rules is proved later.\\

\begin{tabular}{lcc}
(a) & $\qquad$ & $\vcenter{
\infer[\qquad \mathrm{if} ~ U ~ \mathrm{occurs ~ in} ~ \eq ]{\{U =_{}^? V\} \cup \, [V/U](\eq) }
      { \{U =_{}^? V\} ~ \uplus ~ \eq }
}
$\\[+30pt]
(b) & & $\vcenter{
\infer{\eq ~ \uplus ~ \{ U =_{}^? V * W, \; V =_{}^? X, \; W =_{}^? Y \}}
{\eq ~ \uplus ~ \{ U =_{}^? V * W, \; U =_{}^? X * Y \}}
}
$\\[+30pt]
(c) & & $\vcenter{
\infer{\eq ~ \uplus ~ \{ U =_{}^? V \circledast W, \; V =_{}^? X, \; W =_{}^? Y \}}
{\eq ~ \uplus ~ \{ U =_{}^? V \circledast W, \; U =_{}^? X \circledast Y \}}
}
$\\[+30pt]
(d) & & $\vcenter{
\infer{\eq ~ \uplus ~ \{ U =_{}^? exp(V, W), \; V =_{}^? X, \; W =_{}^? Y \}}
{\eq ~ \uplus ~ \{ U =_{}^? exp(V, W), \; U =_{}^? exp(X, Y) \}}
}
$\\[+30pt]
(e) & & $\vcenter{
\infer{\eq ~ \cup ~ \{ U =_{}^? g(V), \; V =_{}^? W \}}
{\eq ~ \uplus ~ \{ U =_{}^? g(V), \; U =_{}^? g(W) \}}
}
$\\[+30pt]
(f) & & $\vcenter{
\infer{\eq ~ \cup ~ \{ U =_{}^? g(X), \; V =_{}^? g(V'), \; X =_{}^? V' \circledast W  \}}
{\eq ~ \uplus ~ \{ U =_{}^? exp(V, W), \; U =_{}^? g(X) \}}
}
$\\[+30pt]
(g) & & $\vcenter{
\infer{\eq ~ \cup ~ \{ U =_{}^? X * Y, \; V =_{}^? V_1^{} * V_2^{}, \; X =_{}^? exp(V_1^{}, W), \; Y =_{}^? exp(V_2^{}, W) \}}
{\eq ~ \uplus ~ \{ U =_{}^? exp(V, W), \; U =_{}^? X * Y \}}
}
$
\end{tabular}

\vspace{0.3in}

\vspace*{0.8em}
The variable $V'$ in rule~(f) is a fresh variable. Similarly
$V_1, V_2$ in rule~(g) are fresh
variables. The symbol $\uplus$ stands for
{\em disjoint union}.
Furthermore, rules~(f) and (g) are applied only when the
other rules cannot be applied. The variable $U$ in rule~(f)
(resp.\ rule~(g)) is called an {\em (f)-peak\/} ({\em (g)-peak.\/}\\

A set of equations is said to be {\em reduced\/} if none of the inference
rules (a) thru (e) are applicable. Thus only rules~(f) and~(g) are applicable
to a reduced system. We define 
relations $\longrightarrow_f^{}$ and $\longrightarrow_g^{}$
between reduced sets
of equations as follows: for reduced sets of equations $S_1$ and $S_2$, 
$S_1 \longrightarrow_f^{} S_2$ (resp., $S_1 \longrightarrow_g^{} S_2$)
if and only if $S_2$ can be obtained from $S_1$ by applying rule~(f) (resp., rule~(g))
{\em once\/} and then eagerly applying rules (a) thru (e). Clearly, rule~(f) decreases the number of $exp$ symbols. But~(g) introduces new $exp$ symbols. Thus termination of the algorithm is not obvious. For simplicity, we assume that the equations deleted while applying the inference are actually put into
``cold storage'' by a marking strategy.\\

Before proceeding we will need to define several relations over the variables in terms of equations
both marked and unmarked. These will be needed later in this paper:

\begin{itemize}

\item[$\bullet$] 
$U ~ \succ_b^{} V$ iff there is an equation $U =_{}^? exp(V, W)$.

\item[$\bullet$] 
$U ~ \succ_e^{} W$ iff there is an equation $U =_{}^? exp(V, W)$.

\item[$\bullet$] 
$U ~ \succ_{l_*}^{} V$ iff there is an equation $U =_{}^? V * W$.
Likewise, $U ~ \succ_{l_\circledast}^{} V$ iff $U =_{}^? V \circledast W$.

\item[$\bullet$] 
$U ~ \succ_{r_*}^{} W$ iff there is an equation $U =_{}^? V * W$.
Likewise, $U ~ \succ_{r_\circledast}^{} W$ iff $U =_{}^? V \circledast W$.

\item[$\bullet$] $U ~ \succ_m^{} ~ V$ iff $U ~ \succ_{l_*}^{} V$ or
$U ~ \succ_{r_*}^{} V$.

\item[$\bullet$] $U ~ \succ_g^{} ~ V$ iff there is an equation $U =_{}^? g(V)$.

\item[$\bullet$] $U ~ \succ ~ V$ iff there is an equation $U =_{}^? t$ such that
$t$ is a non-variable term that contains $V$.


\end{itemize}

Clearly all other relations are sub-relations of $\succ$.
For a relation $p$, let $p_{}^+$ denote its
transitive closure.
Let $\sim$ be the reflexive, symmetric, transitive closure of $\succ_b^{}$.\\

\begin{figure}[h] 
  \centering
   \input{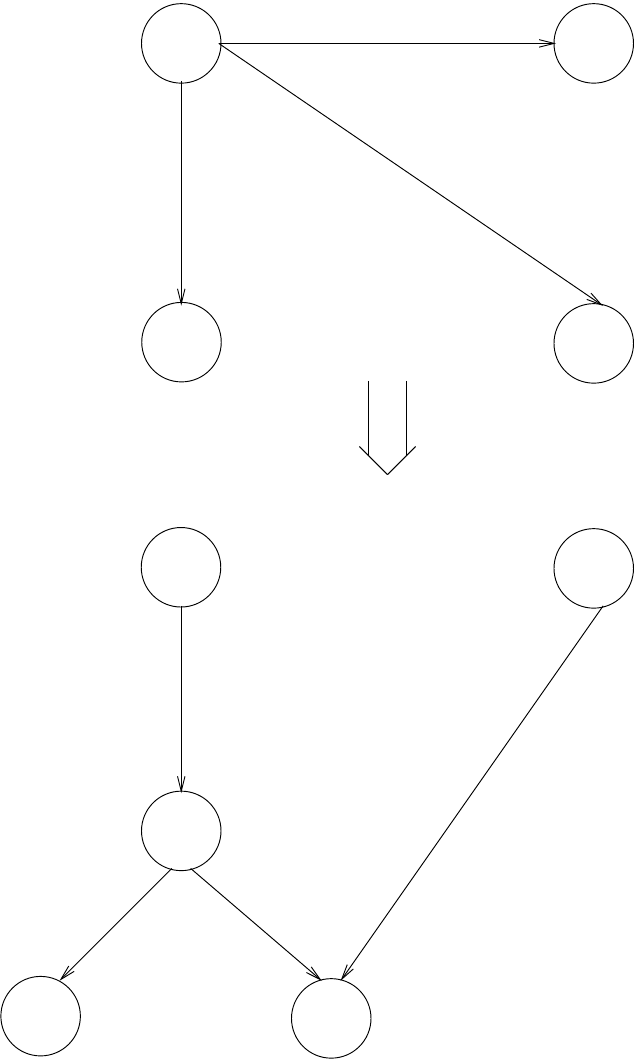}
  \caption{Rule (f)} \label{fig1}
\end{figure}

\begin{figure}[h] 
  \centering
   \input{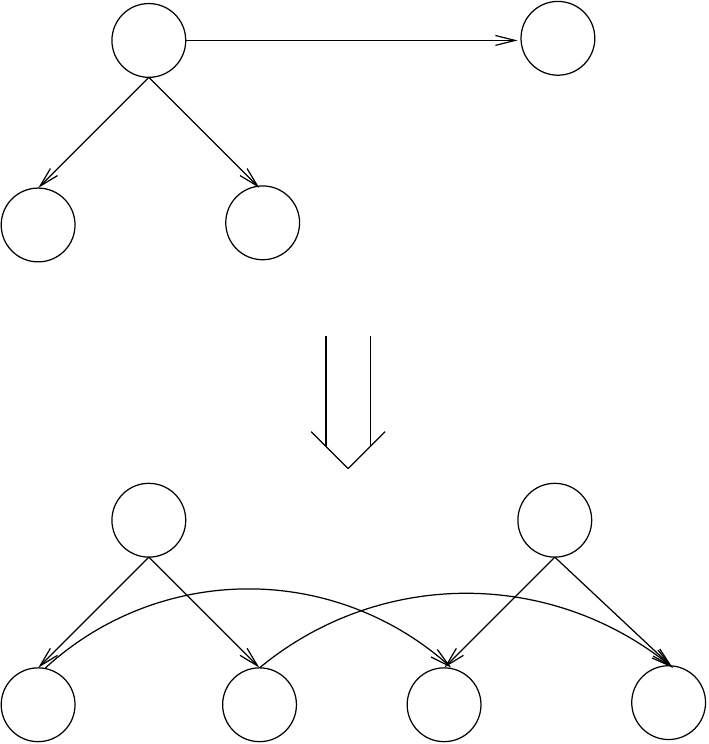}
  \caption{Rule (g): relevant parts} \label{fig2}
\end{figure}

We can also view these relations in terms of graphs, where the
nodes are the variables and the edges correspond to the various
relations between them\footnote{This method is developed by 
Tiden and Arnborg in~\cite{TidenArnborg87}.}. These graphs will be 
useful in checking for failure conditions during unification. 
Figure~\ref{fig1} and Figure~\ref{fig2} are example graphs and 
the resulting transformation after applying an inference rule.

\section{Failure Conditions}
Detection of failure involves several cases. Some cases are caused by 
function clashes and can be detected using the following rules:\\

\begin{tabular}{lcc}
(F1) & & $\vcenter{
\infer{FAIL}
{\eq ~ \uplus ~ \{ U =_{}^? exp(V, W), \; U =_{}^? X \circledast Y \}}
}
$\\[+30pt]
(F2) & & $\vcenter{
\infer{FAIL}
{\eq ~ \uplus ~ \{ U =_{}^? g(V), \; U =_{}^? X \circledast Y \}}
}
$\\[+30pt]
(F3) & & $\vcenter{
\infer{FAIL}
{\eq ~ \uplus ~ \{ U =_{}^? g(V), \; U =_{}^? X * Y \}}
}
$\\[+30pt]
(F4) & & $\vcenter{
\infer{FAIL}
{\eq ~ \uplus ~ \{ U =_{}^? V \circledast W, \; U =_{}^? X * Y \}}
}
$
\end{tabular}

\medskip 
Two other failure cases must be addressed. The first is similar to the
``occur check'' condition in standard unification. The second is a
special case when infinite applications of a rule can happen.  Here we
use congruence classes over the ground terms, i.e. if $t_1$ and $t_2$
are ground terms and $t_1=t_2$ then they are in the same class.
\begin{Lemma} \label{occurcheck}
 Every congruence class over the ground terms is finite. Hence, a term cannot be equivalent to a 
proper-subterm of it. 
\end{Lemma}
\begin{proof}
 The fact that the congruence classes are finite is due to the 
initial system of equations. If a term was 
equivalent to a proper-subterm this would create infinite 
congruence classes by allowing continual replacement of the subterm.  
\end{proof}

\begin{Lemma} \label{claim1}
If there is a variable $X$ such that 
$X \succ_{}^+ X$ then there is no solution.
\end{Lemma}
\begin{proof}
Follows from Lemma~\ref{occurcheck}, this indicates the attempt to 
unify terms in which one is a proper-subterm of the other, 
resulting in an 
occur check failure.
\end{proof}
Next we need to identify cycles between the equivelant classes.
\begin{Lemma} \label{claim2}
If there are two variables $X$ and $Y$ such that
$X \succ_m^{} Y$ and $Y \; (\sim \cup \succ_m^{})_{}^+ \; X$,
then there is no solution.
\end{Lemma}
\begin{proof}
We consider the reduction that
follows when the $exp$ and $\circledast$ functions are
interpreted as a projections onto the first argument.
The reduction will enable a
simpler proof of the result.

\begin{definition} \label{reduction}
 Let $exp$ and $\circledast$ be interpreted as a projection onto the first argument. We define the term $\hat{t}_i$ for any term $t_i$ such that
if $t_i=exp \left( t_{i1}, t_{i2} \right)  $ then
$\hat{t}_i=\hat{t}_{i1}$. Also, if $t_i = t_{i1} \circledast t_{i2} $ then $\hat{t}_i = \hat{t}_{i1}$.
\end{definition}
Consider the Lemma under the interpretation of
Definition~\ref{reduction}. Then any variables related along a
$\sim$ edge will become equivalent. Now consider paths along
$\succ_m$ edges from equivalent classes formed from $\sim$. By
definition there is at least one $\succ_m$ edge from $X$ to
$Y$. We then proceed by induction on the length of the $\succ_m$
path. If no additional $\succ_m$ edges exist we have failure due
to $X \succ_m Y$ and $Y \sim X$ ($X=Y$). Now we can see that
adding $\sim$ edges will not effect the unification of the
system. we then can assume that we have a cycle of $k \left( 0
\leq k \right) $ $\succ_m$ edges that do not form a unifiable
system. That is, a cycle of the form $E_1 \succ_m E_2 \succ_m
\cdots \succ_m E_{k+1}$, where each $E_i$ is an equivalence
class, $Y \in E_1$, $X \in E_{k+1}$ and $X \succ_m Y$. Then
because adding another $\succ_m$ edge would only move $X$ into a
lower class we can see the cycle is not unifiable.
\end{proof}

\begin{Lemma} \label{claim3}
If $\{ U =_{}^? X * Y, \; V =_{}^? g(Z) \} ~
\subset ~ \eq$ and $U \sim V$ then there is no solution.
\end{Lemma}
\begin{proof}
 Because of the bi-directional nature of $\sim$ we prove both directions. 

First: let $u=x*y$, $v=g \left( z \right) $ and $u \succ_b^+ v$. If $u
\succ_b v$ we must unify the equations $x*y$ and $exp \left( v,w
\right) $ but this immediately leads to a function clash due to the
need to unify $v=g \left( z \right) $ and $v=v_1 * v_2$. We can see
that for any path along $\succ_b^+$ we can continue to move the $*$
along the path until eventually we will need to unify $v=g(z)$ and
$v=v_1 * v_2$.\\

Second: Let $u=x*y$, $v=g(z)$, and $v \succ_b^+u$. Just as in the
first direction then we can move the $g$ function along the $\succ_b$
path eventually we will be required to unify $u=x*y$ and $u=g(v^{'})$,
a function clash.
\end{proof}

\section{Unification Algorithm}

First we need a method for detecting ``occur check'' failure
conditions. To accomplish this, we use the methods developed in Tiden
and Arnborg~\cite{TidenArnborg87}, building two special graphs to
check for failure conditions.

\begin{definition} \label{D}
Let $D$ be a graph defined on a reduced system of equations. The
nodes in the graph correspond to variables in the system. The edges
correspond to the parameters of each equation type. See Figure~\ref{fig1}.
\end{definition}

\begin{Lemma} \label{Lemma1}
If there exists a cycle in $D$, the set of equations 
represented by $D$ is not unifiable.
\end{Lemma}
\begin{proof}
 Directly from Lemma~\ref{claim1}.
\end{proof}

\begin{figure}[h] 
  \centering
   \input{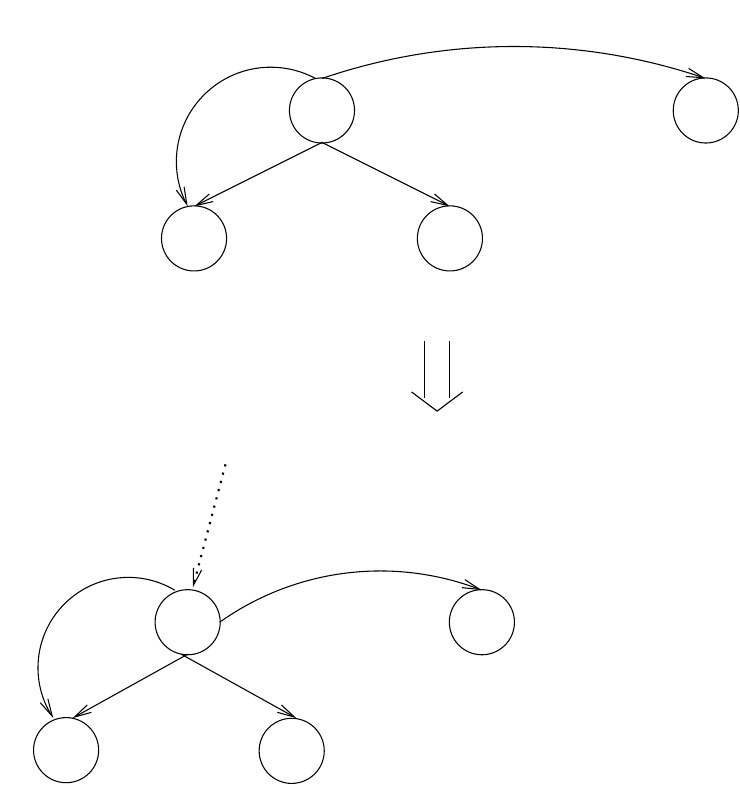}
  \caption{$U=exp \left( X, W \right) $, $U= X * Y$} \label{inf}
\end{figure}

\noindent
We will also need to detect cases requiring an infinite unifier. An
example of this is the set of equations 
comprising $U \ueq exp\left( X, W \right)$, and
$U \ueq X * Y$. This example (g)-peak would cause a new (g)-peak creation
after each application of Rule~(g) (See Figure~\ref{inf}). We will use
a propagation graph $P$ to check for these conditions.

\begin{definition} \label{P}
Let $P$ be a directed simple graph defined on a set of equations as follows:
Each vertex in $P$ is a $\sim$-equivalence class.
There is an edge between the vertex containing $v$ to the vertex 
containing $w$ in $P$, if there is a $\succ_{m}$  labeled 
edge from $v$ to $w$ in $D$.
\end{definition}

\begin{Lemma} \label{Lemma2}
If there exists a cycle in $P$, the set of equations 
represented by P is not unifiable.
\end{Lemma}
\begin{proof}
\ignore{This result is along the same lines as that in Tiden and Arnborg~\cite{TidenArnborg87} regarding
their `sum propagation graph', where it is
shown that the $S_{1} \Rightarrow_s^{} S_{2} \Rightarrow_s^{} \cdots$
infinite sequence requires an infinite unifier. The propagation graph
here is a simple modification of their presentation using $\sim$. This
translates into cycles in $P$ corresponding to infinite chains along the
$\succ_b$ edges and thus nonunifiability.}
Follows from Lemma~\ref{claim2}.
\end{proof}

We now give a general unification algorithm for unification 
modulo the partial theory of exponentiation.

\begin{algorithm}[H]
\caption{Unification modulo partial exponentiation}
\label{alg1}
\begin{algorithmic}
\REQUIRE $EQ$, the set of equations
\WHILE{An inference rule can be applied}
\item Build graphs D and P; if a cycle is found exit with failure.
\item If any of rules (F1) through (F4) apply exit with failure.
\item Eagerly apply rule (a).
\item Eagerly apply rules (b) through (e). 
\item Apply rules (f) and (g) if possible. 
\ENDWHILE
\end{algorithmic}
\end{algorithm}

\begin{figure}[H] 
  \centering
   \input{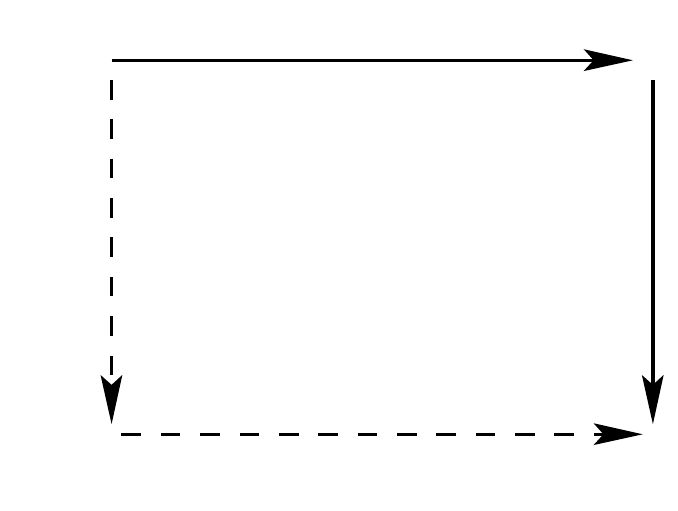}
  \caption{Rule (f) commutes with Rule (g)} \label{comm}
\end{figure}

\begin{Lemma} \label{lemma1}
 Rule~(f) commutes with rule~(g). (See Figure~\ref{comm})

\end{Lemma}
\begin{proof}
No variable can be an (f)-peak and a (g)-peak at the same time
because this would cause failure. Thus, application of rule~(g) first 
does not affect
the applicability of rule~(f).\\
\end{proof}

\begin{theorem} \label{halt}
Algorithm~\ref{alg1} always terminates.
\end{theorem}
\begin{proof}
If a failure condition or cycle in one of the graphs is found, 
Algorithm~\ref{alg1} will clearly halt. Assume none of these 
conditions occur. Then some observations can be made: Every
$\sim$-congruence class has to have a unique sink (wrt~$\succ_b^{}$). 
Also, applying rule~(g) does not increase the number of congruence classes ---
the new variables $V_1^{}$ and $V_2^{}$ are $\sim$-equivalent to 
$X$ and $Y$ respectively.
Now $\succ$ can be used to define a 
well-founded partial order
on the {\em $\sim$-congruence classes.\/} Thus the new $exp$ equations created
in rule~(g) are on congruence classes \underline{lower} than the earlier one.
Applications of rule~(g) will thus always terminate 
under the above assumptions. 
Since rule~(f) can potentially increase the number of congruence classes, we
need Lemma~\ref{lemma1}. Since one cannot get an 
infinite sequence of (g)-steps or (f)-steps, the algorithm
terminates.
\end{proof}

\section{Undecidability of unification of partial exponentiation with two Abelian group operators}
Let us now consider the expanded theory where both $*$ and $\circledast$ are 
Abelian group 
operations. That is, we let $*$ represent 
multiplication modulo a 
prime $p$ and $\circledast$ represent  
multiplication modulo $p-1$. We denote this equational
theory as $\mathcal{E}_1$ and the resulting AC-convergent system 
as $\mathcal{R}_1$:\\[-5pt]

\begin{minipage}[t]{0.5\linewidth}
\begin{footnotesize}
\begin{enumerate}
\item[] $X * X^{-1} \rightarrow 1$
\item[] $X * 1 \rightarrow X$
\item[] $\left( X * Y \right)^{-1} \rightarrow X^{-1} * Y^{-1} $
\item[] $ \left( \left( Z \right)^{-1} \right)^{-1} \rightarrow Z$
\item[] $ 1^{-1} \rightarrow 1 $
\item[] $ X \circledast 1 \rightarrow X$
\item[] $ X \circledast i\left( X \right) \rightarrow 1  $
\item[] $ i\left( i\left( X \right)  \right)  \rightarrow X $
\item[] $ i\left( X \circledast Y \right) \rightarrow i\left( X \right) \circledast i\left( Y \right)  $
\end{enumerate}
\end{footnotesize}
\end{minipage} \
\begin{minipage}[t]{0.5\linewidth}
\begin{footnotesize}
\begin{enumerate}
\item[] $\qquad$
\item[] $ exp \left( X , 1 \right) \rightarrow x$
\item[] $exp \left( 1, Z \right) \rightarrow 1$
\item[] $exp \left( Z^{-1}, X \right) \rightarrow \left( exp \left( Z, X \right) \right)^{-1} $
\item[] $exp \left( g \left( X \right) , Y \right) \rightarrow g \left( X \circledast Y \right)  $
\item[] $exp \left( \left( X * Y \right), Z \right) \rightarrow exp \left( X, Z \right) * exp \left( Y, Z \right) $
\end{enumerate}
\end{footnotesize}
\end{minipage}

\vspace{0.1in}

\noindent
where $<\!\! *, {}^{-1}, 1 \!\!>$ forms the first Abelian group and
$<\!\! \circledast, i(), 1 \!\!>$ the second. The unification problem for this
system is undecidable. The proof is by reduction from Hilbert's
$10^{th}$ problem (solvability of polynomial equations over the
integers). It will be shown that multiplication and addition of a
number can be simulated in the above system. We make the assumption
for the first part of the proof that we are allowed the distinct free
constants $b$ and $c$. 
\ignore{Later it is shown that we can remove this
assumption and preserve the undecidability result.} The following proof
is a modification of the proof given in~\cite{Narendran93onthe}.\\
\ignore{
\begin{definition}
 Let $a^{x+y}$ be an abbreviation of $a^x * a^y$. Let $b^i$ abbreviate $\underbrace{ b*b* \ldots *b}_i$. 
\end{definition}
}
\begin{definition}
 Let $\bigcirc_i(u)$ denote 
\begin{itemize}
 \item $ \underbrace{u \circledast u \circledast \ldots \circledast u}_i$, if $i > 0$.
\item $\bigcirc_i(u) = 1$ if $i = 0$ and
\item $ \underbrace{i(u) \circledast i(u) \circledast \ldots \circledast i(u)}_i$, if $i < 0$. 
\end{itemize}

\end{definition}

\vspace{0.1in}
\begin{Lemma} \label{Lemma7}
$g \left( s \right) =_{\mathcal{E}_1}^{} g \left( t\right) 
\Rightarrow s =_{\mathcal{E}_1}^{} t $. 

\end{Lemma}

\vspace{0.1in}
\begin{Lemma}
 For every $m\text{, }n \in \mathbb{Z}$, the equation:\\
\[ x* g\left( \bigcirc_{n}(b) \right) ~ {=_{\mathcal{E}_1}^{}} ~
    exp \left( x, b \right) * g( \bigcirc_{m}(b) ) \]
is solvable.

\end{Lemma}

\vspace{0.1in}
\begin{proof}$ $\\
 \begin{itemize}
\item[(a)] If $n > m$, then  
$x=g\left( \bigcirc_{n-1}(b)\right) * \ldots * g\left(\bigcirc_{m} (b) \right)$ 
is a solution. \\
\item[(b)] If $n < m$, then 
$x=(g\left( \bigcirc_{n}(b)\right) * \ldots * g\left(\bigcirc_{m-1} (b) \right))_{}^{-1}$
is a solution. \\
\item[(c)] If $n=m$, then $x=1$ is a solution.\\[-32pt]
\end{itemize}
\end{proof}

\vspace{0.1in}
\noindent
\begin{Lemma} \label{Lemma8}
Let $b$ be a free constant and $m$ be an integer. Then, every solution 
to \[ x* g\left(y\right) ~ {=_{\mathcal{E}_1}^{}} ~
    exp \left( x, b \right) * g( \bigcirc_{m}(b) ) \] is of one of the 
following forms:
\begin{itemize}
\item[(a)] $n > m$,  $y=\bigcirc_{n}(b)$,
$x=g\left( \bigcirc_{n-1}(b)\right) * \ldots * g\left(\bigcirc_{m} (b) \right)$\\
\item[(b)] $n < m$, $y=\bigcirc_{n}(b)$,
$x=(g\left( \bigcirc_{n}(b)\right) * \ldots * g\left(\bigcirc_{m-1} (b) \right))_{}^{-1}$\\
\end{itemize}
\end{Lemma}

\begin{proof}
The proof is by contradiction. Suppose that there exist
an integer $m$ and terms $t_x^{}$ and $t_y^{}$,
in normal form modulo $ {{\mathcal{R}_1}^{}}$, such 
that \[ t_x^{} * g( t_y^{} ) ~ {=_{\mathcal{E}_1}^{}} ~ exp(t_x^{}, b) * g( \bigcirc_{m}(b) ) \] where
$t_y^{} \neq \bigcirc_{n}(b)$ for any $n$. Without loss of generality
assume also that $t_x^{}$ is a minimal (by size) counterexample, i.e., a minimal
term such that $\exists m \, \exists t_y^{} : \; t_x^{} * g( t_y^{} ) ~ {=_{\mathcal{E}_1}^{}} ~ exp(t_x^{}, b) * g( \bigcirc_{m}(b) )$.

First of all note that since $ {{\mathcal{R}_1}^{}}$ is AC-convergent, it must be 
that \[ t_x^{} * exp(t_x^{-1}, b) * g( t_y^{} ) ~ \rightarrow_{{\mathcal{R}_1}^{}}^! ~
g( \bigcirc_{m}(b) ). \]\\
Then $t_x$ can have two possible forms:\\

\noindent
\underline{Case 1:}  $t_x = g(\bigcirc_{m}(b)) * t_x^{'}$. Then,\\
\begin{eqnarray*}
g(\bigcirc_{m}(b)) * t_x^{'} * g(t_y)  & =_{\mathcal{E}_1}^{} & exp(g(\bigcirc_{m}(b)), b) * exp(t_x^{'} , b) * g(\bigcirc_{m}(b))
~ ~  \qquad \mathrm{and ~ thus}\\
t_x^{'} * g(t_y)  & =_{\mathcal{E}_1}^{} & g(\bigcirc_{m+1}(b)) * exp(t_x^{'}, b) 
\end{eqnarray*}
Thus $t_x^{'}$ is a smaller counterexample.\\

\noindent
\underline{Case 2:}  $t_x = g(\bigcirc_{m-1}(b))^{-1} * t_x^{'}$. Then,\\
\begin{eqnarray*}
g(\bigcirc_{m-1}(b))^{-1} * t_x^{'} * g(t_y)  & =_{\mathcal{E}_1}^{} & exp(g(\bigcirc_{m-1}(b)), b)^{-1} 
* exp(t_x^{'} , b) * g(\bigcirc_{m}(b)) ~ ~  \qquad \mathrm{and ~ thus}\\
t_x^{'} * g(t_y)  & =_{\mathcal{E}_1}^{} & g(\bigcirc_{m-1}(b)) * exp(t_x^{'}, b) 
\end{eqnarray*}
Thus $t_x^{'}$ is a smaller counterexample.
\end{proof}

\noindent
\begin{Lemma} \label{Lemma9}
Let $b$ and $c$ be free constants. Then, the equations
\begin{eqnarray*}
 exp\left( x, c\right) * g\left( \bigcirc_{j}(b) \right) & =_{\mathcal{E}_1}^{} & exp \left( x, b \right) * g\left( u \right)   \\
z * g \left( u \right)  & =_{\mathcal{E}_1}^{} & exp \left( z,c \right) * g(1)
\end{eqnarray*}
force $u$ to be equal to $\bigcirc_{j}(c)$.
\end{Lemma}
\begin{proof}
By Lemma~\ref{Lemma8} the second equation,
$ z * g \left( u \right)   =_{\mathcal{E}_1}^{}  exp \left( z,c \right) * g(1) $,
forces $ u= \bigcirc_{n}(c) $.
Now replacing $b$ with $c$ everywhere in the first equation 
we get \[ exp(x, c) * g(\bigcirc_{j}(c)) = exp(x, c) * g( \bigcirc_{n}(c)). \]
By Lemma~\ref{Lemma7}
$ \bigcirc_{j}(c) = \bigcirc_{n}(c) $ and $j=n$.
\end{proof}

\vspace{0.15in}

\noindent
\begin{Lemma} \label{Lemma10}
Let $b$ and $c$ be free constants. Then the equations:
\begin{eqnarray*}
exp \left( x, \bigcirc_{k}(c) \right) * g \left(  \bigcirc_{j}(b) \right)  & =_{\mathcal{E}_1}^{} & 
exp \left( x, b \right) * g \left( u \right)  \\
z * g \left( u \right) & =_{\mathcal{E}_1}^{} & exp \left( z, c \right) * g(1) 
\end{eqnarray*}
force $u$ to be equal to $\bigcirc_{jk}(c)  $
\end{Lemma}
\begin{proof}
By Lemma~\ref{Lemma8} $u= \bigcirc_{n}(c)$ as before. Now
replacing $b$ by $ \bigcirc_{k}(c)$ we get \[
exp(x, \bigcirc_{k}(c)) * g(\bigcirc_{jk}(c)) = exp(x, \bigcirc_{k}(c)) * g( \bigcirc_{n}(c)). \]
By Lemma~\ref{Lemma7} $\bigcirc_{jk}(c) = \bigcirc_{n}(c)$ and $n = jk$.
\end{proof}

With Lemma~\ref{Lemma10} we can now simulate multiplication 
with the natural numbers. To see how this can be done consider 
$z=x*y$ and let $x=\bigcirc_{i}(b) $ and $y=\bigcirc_{j}(b) $. 
We force $z=\bigcirc_{ij}(b)  $ 
as follows:
\begin{eqnarray*}
exp \left( w_1, c \right) * g \left(  \bigcirc_{i}(b) \right) & =_{\mathcal{E}_1}^{} & exp \left( w_1, b \right) * 
g \left(  x_2 \right) ~ ~  \qquad \mathrm{and}\\
w_2 * g \left( x_2 \right) & =_{\mathcal{E}_1}^{} & exp \left( w_2, c \right) * g(1) 
\end{eqnarray*}
force $x_2= \bigcirc_{i}(c) $ by Lemma~\ref{Lemma9}.
\begin{eqnarray*}
exp \left( w_3, x_2 \right) * g \left(  \bigcirc_{j}(b) \right) & =_{\mathcal{E}_1}^{} & 
exp \left( w_3, b \right) * g \left( z_2 \right) ~ ~ \qquad \mathrm{and}\\
w_4 * g \left( z_2 \right) & =_{\mathcal{E}_1}^{} & exp \left( w_4, c \right) * g(1) 
\end{eqnarray*}
force $z_2 = \bigcirc_{ij}(c) $ by Lemma~\ref{Lemma10}.
Finally we copy $z_2$ to $z$ with the equation
\begin{eqnarray*}
exp \left( w_5, c \right) * g \left( z \right) & =_{\mathcal{E}_1}^{} & exp \left( w_5, b \right) * g \left( z_2 \right).
\end{eqnarray*}

\begin{Lemma} \label{add}
Addition of natural numbers 
can be simulated in $\mathcal{E}_1$.
\end{Lemma}
\begin{proof}
Let $x= \bigcirc_{i}(b) $ and $y= \bigcirc_{j}(b) $, 
where $b$ is a free constant. Then
$x \circledast y ~ =_{\mathcal{E}_1}^{} ~ \bigcirc_{i+j} (b)$
\end{proof}

\begin{theorem}
Unification over  $\mathcal{E}_1$ with free constants is undecidable.
\end{theorem}
\begin{proof}
Following the above outline a unification problem can be 
constructed that simulates a system of diophantine equations.
\end{proof}


\section{Extension and Limitations}

In this paper we examined a partial theory of exponentiation, a
critical component in several cryptographic protocols.  Many of
the protocols based on modular exponentiation also contain additional
algebraic properties and axioms that could correspond to
extensions of this partial exponentiation theory. Therefore, an
important question that naturally arises is, how far we can
extend the theory and maintain decidability. Unfortunately,
additional extensions can quickly result in undecidable
unification problems. This was demonstrated when the operations
of $\circledast$ and $*$ were allowed to form abelian
groups. Therefore, ideally, extensions should maintain decidability
while adding additional axioms useful in modeling additional
cryptographic protocols. We are currently examining two different
possible extensions.  The first is allowing just one of either
the $\circledast$ or $*$ operations to be abelian.  The second is
extending the axiom set to include additional algebraic operators
such as {\em modular addition.\/}  Several other papers,
including~\cite{Meadows02, KNW03b, KNW03a, Kapur05}, have also
considered the unification problem for equational systems that
contain some type of exponentiation. For convenience, we give a
condensed overview of a selection of these results in
Table~\ref{table0}.

\begin{table}[H]
\begin{center}
    \begin{tabular}{|l|p{10cm}|p{5cm}|}
    \hline
 Ref   & Equational Theory & Unification Problem: Results    \\ \hline
    ~\cite{Meadows02} & Abelian group with the axioms $exp(x,1)=1$ and 
$exp(exp(x,y)z) = exp(x, y*z)$ & NP-complete   \\ \hline
    ~\cite{KNW03b} & Two theories, denoted $\E_1$ and $\E_2$. 
$\E_1$ consists of an abelian group with operator, $\cdot$, and a monoid with operator $\circ$ 
with the addition of the axioms:
$x^{1} = x$, $1^{x} = 1$, $(x \cdot y)^{z} = (x^z) \cdot (y^z)$, and $(x^{y})^{z} = x^{y \circ z}$.
$\E_2$ adds the axiom $x \circ i(x)=1$, $i(x)$ being the inverse, to the theory $\E_1$.   
& Undecidable for both $\E_1$ and $\E_2$  \\ \hline
    ~\cite{KNW03a} & Two main results: Theory $\E_3$ consists of an abelian group for operator $\cdot$ 
along with the axioms, $x^{1} = x$, $1^{x} = 1$, and $(x \cdot y)^{z} = (x^z) \cdot (y^z)$. 
Theory $\E_{4}$ consists of $\E_3$ with the addition of a monoid operator $\circ$ and the 
axiom $(x^{y})^{z} = x^{y \circ z}$. 
 & $\E_3$ is decidable and $\E_{4}$ is undecidable.  \\ \hline
    ~\cite{Kapur05} & Two theories, denoted $\E$ and $\E_0$. $\E$ consists
of an abelian group with operator, $\cdot$, and the axioms 
$x^{1} = x$, $1^{x} = 1$, $(x \cdot y)^{z} = (x^z) \cdot (y^z)$, and $(x^{y})^{z} = x^{y \cdot z}$.
$\E_0$ is the same as $\E$ but the axiom $(x^{y})^{z} = x^{y \cdot z}$ is replaced with
the axiom $x^{y^z} = x^{z^y}$ 
& $\E$ is undecidable and $\E_0$ is decidable.  \\ \hline
    \end{tabular}
\caption{Results for E-unification with exponentiation.}
  \label{table0}
\end{center}
\end{table}   
Most of these results are of high complexity. Therefore, we are also exploring heuristic
methods of implementation to enable their integration into the automated protocol analysis system 
Maude-NPA~\cite{EscobarMM-fosad}.  

\bibliographystyle{eptcs} 
\bibliography{partialxp} 
\nocite{BaaderSnyder,KNW03a}

\end{document}